\documentstyle[preprint,aps]{revtex}
\tightenlines

\def\singlespace {\smallskipamount=3.75pt plus1pt minus1pt
                  \medskipamount=7.5pt plus2pt minus2pt
                  \bigskipamount=15pt plus4pt minus4pt
                  \normalbaselineskip=12pt plus0pt minus0pt
                  \normallineskip=1pt
                  \normallineskiplimit=0pt
                  \jot=3.75pt
                  {\def\smallskip {\vskip\smallskipamount}}
                  {\def\medskip   {\vskip\medskipamount}}
                  {\def\bigskip   {\vskip\bigskipamount}}
                  {\setbox\strutbox=\hbox{\vrule
                    height10.5pt depth4.5pt width 0pt}}
                  \parskip 7.5pt
                  \normalbaselines}
\def\middlespace {\smallskipamount=5.625pt plus1.5pt minus1.5pt
                  \medskipamount=11.25pt plus3pt minus3pt
                  \bigskipamount=22.5pt plus6pt minus6pt
                  \normalbaselineskip=22.5pt plus0pt minus0pt
                  \normallineskip=1pt
                  \normallineskiplimit=0pt
                  \jot=5.625pt
                  {\def\smallskip {\vskip\smallskipamount}}
                  {\def\medskip   {\vskip\medskipamount}}
                  {\def\bigskip   {\vskip\bigskipamount}}
                  {\setbox\strutbox=\hbox{\vrule
                    height15.75pt depth6.75pt width 0pt}}
                  \parskip 11.25pt
                  \normalbaselines}
\def\doublespace {\smallskipamount=7.5pt plus2pt minus2pt
                  \medskipamount=15pt plus4pt minus4pt
                  \bigskipamount=30pt plus8pt minus8pt
                  \normalbaselineskip=30pt plus0pt minus0pt
                  \normallineskip=2pt
                  \normallineskiplimit=0pt
                  \jot=7.5pt
                  {\def\smallskip {\vskip\smallskipamount}}
                  {\def\medskip   {\vskip\medskipamount}}
                  {\def\bigskip   {\vskip\bigskipamount}}
                  {\setbox\strutbox=\hbox{\vrule
                    height21.0pt depth9.0pt width 0pt}}
                  \parskip 15.0pt
                  \normalbaselines}
\def\al{\alpha}

\def\th{\theta}

\def\si{\sigma}

\def\ph{\phi}

\def\Ph{\Phi}

\def\cF{{\cal F}}

\def\frac#1#2{\textstyle{{{#1} \over {#2}}}}

\def\prt{\partial}

\def\lsim{\mathrel{\rlap{\lower4pt\hbox{\hskip1pt$\sim$}}
    \raise1pt\hbox{$<$}}}
\def\gsim{\mathrel{\rlap{\lower4pt\hbox{\hskip1pt$\sim$}}
    \raise1pt\hbox{$>$}}}

\newcommand{\beq}{\begin{equation}}
\newcommand{\eeq}{\end{equation}}
\newcommand{\bea}{\begin{eqnarray}}
\newcommand{\eea}{\end{eqnarray}}

\tightenlines
\begin{document}
\preprint{
\hfill$\vcenter{\hbox{\bf IUHET-479} \hbox{February 
             2004}}$  }

\title{\vspace*{.75in}
Spacetime Symmetry 
Violation\footnote{\uppercase{T}alk presented 
at {\it \uppercase{SUSY} 2003:
\uppercase{S}upersymmetry in the \uppercase{D}esert}\/, 
held at the \uppercase{U}niversity of \uppercase{A}rizona,
\uppercase{T}ucson, \uppercase{AZ}, \uppercase{J}une 5-10, 2003.
\uppercase{T}o appear in the \uppercase{P}roceedings.}}

\author{M. S. Berger
\footnote{Electronic address:
berger@indiana.edu}}

\address{
Physics Department, Indiana University, Bloomington, IN 47405, USA}

\maketitle

\thispagestyle{empty}

\begin{abstract}
Supersymmetric models with Lorentz violation can be formulated in 
superspace. Two theories based on the Wess-Zumino model are discussed.
A compactification of superspace can be employed to understand the 
chiral superfield that arises in the models.
\end{abstract}

\newpage

\section{Introduction}
 
Spacetime symmetries have played an important part in our understanding of 
fundamental physics. The discovery of special relativity as well as the 
Lorentz symmetry underlying it was followed 
by the proposal for a larger spacetime 
symmetry, namely supersymmetry. In this case
experimental observations require that if 
spacetime supersymmetry is relevant for describing particle physics, then it 
must be broken. More recently there has been extended discussion of 
extra dimensions. If the extra dimensions are compactified, then there is 
necessarily a violation of the extended Lorentz symmetry that applies to the
extra dimensions. This history should encourage us to consider possible 
connections between these various broken spacetime symmetries (and the various
scales involved in the breaking) as well as consider the possibility that the 
four-dimensional Lorentz symmetry is itself violated even though there is 
at present
no experimental evidence for it. In the following we discuss supersymmetric 
models based on the Wess-Zumino model that contain Lorentz violation.

\section{Superspace Transformations}

The Wess-Zumino model\cite{Wess:tw}
 can be formulated in terms of differential operators that
act on the superfields defined over a superspace of coordinates
\bea
&&z^M=(x^\mu,\th ^\al,\bar\th _{\dot\al})\;.
\eea 
where $x^\mu$ are commuting spacetime coordinates and 
$\th ^\al$ and $\bar\th _{\dot\al}$ are anticommuting
two-component Weyl spinors.
Let
\bea
&&X\equiv (\th \sigma ^\mu \bar\th)\prt _\mu\;, 
\label{Xop}
\eea
so that
\bea
&&U_x \equiv e^{iX}=1+i(\th \sigma ^\mu \bar\th)\prt _\mu -{1\over 4}(\th \th)
(\bar\th \bar\th)\Box \;. 
\eea
Application of $U_x$
to a superfield ${\mathcal S}$ produces a coordinate shift 
$x^\mu \to y^\mu=x^\mu+i\th \si ^\mu \bar \th$,
\bea
&&U_x{\mathcal S}(x,\th,\bar\th)={\mathcal S}(y,\th,\bar\th)\;.
\label{shift}
\eea
A {\it chiral} superfield is a function of 
$y^\mu$ and $\th$,
\bea 
\Ph(x,\th,\bar\th)&=&\ph(y)+\sqrt{2}\th \psi(y)+(\th \th)\cF(y)\;,
\end{eqnarray}
so it can be expressed in terms of 
a superfield $\Psi$ in the following manner
$\Ph(x,\th,\bar\th)=U_x\Psi(x,\th)$ where
$\Psi$ depending on only $x^\mu$ and $\th$ and not $\bar{\th}$. 
The Wess-Zumino model can be expressed as
\bea
&&\int d^4\th \left [U_x^*\Psi(x,\bar\th)^*\right ]\left [U_x\Psi(x,\th)
\right ]
+ \int d^2\th \left [ 
{1\over 2}m\Ph^2 +{1\over 3}g\Ph^3 
+h.c.\right ]\;.
\eea

The Lorentz-violating extensions\cite{Berger:2001rm} of the Wess-Zumino
model can be understood in an analogous way as Lorentz-violating
transformations on the superfields similar to the one in Eqn.~(\ref{shift}).
Considering the derivative operator in Eqn.~(\ref{Xop}), define
\bea
&&Y\equiv k_{\mu\nu}(\th \sigma ^\mu \bar\th)\prt ^\nu\;, \\
&&K\equiv k_\mu(\th \sigma ^\mu \bar\th)\;,
\eea
so that
\bea
&&U_y \equiv e^{iY}=1+ik_{\mu\nu}(\th \sigma ^\mu \bar\th)\prt^\nu
 -{1\over 4}k_{\mu \nu}k^{\mu \rho}(\th \th)
(\bar\th \bar\th)\prt ^\nu \prt _\rho\;, \\
&&T_k \equiv e^{-K}=1-k_\mu(\th \sigma ^\mu \bar\th)+{k^2\over 4}(\th \th)
(\bar\th \bar\th)\;.
\eea
Terms necessarily appear that are quadratic in the 
Lorentz-violating coefficients $k_{\mu\nu}$ and $k_\mu$.
Since $Y$, like $X$, is a derivative operator, the action of $U_y$
on a superfield ${\mathcal S}$
is a coordinate shift. On the other hand, $T_k$ is not a derivative operator
and its action does not shift the spacetime coordinate. Consequently the 
application of these operators is a generalization of the conventional 
coordinate shift $x^\mu \to y^\mu$ usually associated with chiral superfields. 
The following properties are satisfied: $U_x^*=U_x^{-1}$, $U_y^*=U_y^{-1}$ 
and $T_k^*=T_k$.

A first supersymmetric model with Lorentz-violating terms 
can be expressed in terms of a new superfield\cite{Berger:2003ay},
\bea
\Ph_y(x,\th,\bar\th)&=&U_yU_x\Psi(x,\th)\;. 
\eea
Applying $U_y$ to the chiral and antichiral superfields merely effects the
substitution $\prt _\mu \to \prt _\mu +k_{\mu\nu}\prt ^\nu$. The chiral 
superfield $\Ph_y$ is a function of the variables 
$x_+^\mu=y^\mu ++ ik^{\mu \nu}\th 
\si_\nu\bar\th=x^\mu + i\th \si ^\mu \bar\th + ik^{\mu \nu}\th 
\si_\nu\bar\th$
and $\theta$ analogous to how, in the usual case, $\Ph$ is a function 
of the variables $y^\mu$ and $\theta$. The 
Lagrangian is given in terms of integrals over superspace 
\bea
&&\int d^4\th \Ph_y^*\Ph_y+ \int d^2\th \left [ 
{1\over 2}m\Ph_y^2 +{1\over 3}g\Ph_y^3 
+h.c.\right ]\nonumber \\
&&=\int d^4\th \left [U_y^*\Ph^*\right ]
\left [U_y\Ph\right ]+ \int d^2\th \left [ 
{1\over 2}m\Ph^2 +{1\over 3}g\Ph^3 
+h.c.\right ]\;.
\label{superspace2}
\eea

A superfield appropriate for a second supersymmetric model with Lorentz 
violation has the form
\bea
\Ph_k(x,\th,\bar\th)&=&T_kU_x\Psi(x,\th)\;,
\eea
The superspace integral
\bea
&&\int d^4\th \Ph_k^*\Ph_k=\int d^4\th \Ph^* e^{-2K}\Ph \;,
\label{proj}
\eea
describes a CPT-violating model. The 
$(\th \th)(\bar\th \bar\th)$ component of $\Ph_k^*\Ph_k$
transforms into a total derivative.

\section{Supermanifolds}

It is clear that compactification of spacetime results in violations of the 
Lorentz symmetry\footnote{An elementary example of the physical implications
of the global structure of spacetime is a modified twin paradox.
It is only necessary to consider one space and one time dimension.
In this case, the twins are on a cylinder $R\times S_1$ and the
compactification picks out a preferred frame\cite{twins}.}.
Compactification of extra dimensions (those beyond the 
conventional four) inevitably leads to violations of the Lorentz group 
that is extended to those extra dimensions.
Furthermore models of the Scherk-Schwarz 
variety\cite{Scherk:1978ta,Scherk:1979zr,Cremmer:1979uq} result in 
broken supersymmetry in four dimensions from
compactification of extra dimensions.

In complexified superspace\cite{Buchbinder:qv} one can understand the 
chiral superfields as those defined after a suitable 
compactification of a 
supermanifold\cite{Rabin:1984rm}. Under the transformation
\begin{eqnarray}
&&(x^\mu,\theta,\bar{\theta})\to (x^\mu+i\theta \sigma^\mu \bar{\eta},
\theta,\bar{\theta}-\bar{\eta})\;,
\label{trans}
\end{eqnarray}
a chiral superfield is invariant. If one mods out by the discrete subgroup
of transformations where the components of $\eta$ are complex Grassman integers
$m+in$ for integers $m$ and $n$, then 
only those superfields which are invariant under the 
transformation in Eqn.~(\ref{trans}) are defined on the quotient 
space\cite{Rabin:1984rm} since a superfield must be constant along 
a compact direction. It is straightforward to understand 
a Lorentz-violating extension of the Wess-Zumino model as a compactification
that does not respect the Lorentz symmetry. The relevant 
transformation on superspace would be
\begin{eqnarray}
&&(x^\mu,\theta,\bar{\theta})\to (x^\mu+i\theta \sigma^\mu \bar{\eta}
+ik^{\mu\nu}\theta \sigma_\nu \bar{\eta},
\theta,\bar{\theta}-\bar{\eta})\;.
\end{eqnarray}
Forming the quotient space in the same way results in a chiral superfield
that depends on the invariant variables $x_+^\mu$ and $\theta$. Clearly the 
hope is to link Lorentz violation with supersymmetry breaking in this way.


\section{Conclusions}

The Wess-Zumino model and two Lorentz-violating extensions of it
can be described in terms of transformations on  
superfields and projections arising from superspace integrals.
A geometric interpretation is available by considering a compactification 
of a complexified superspace.


\vspace{0.5cm}

\section*{Acknowledgements}

This work was supported in part by the U.S.
Department of Energy under Grant No.~DE-FG02-91ER40661.


\end{document}